\documentclass[11pt]{article}
\usepackage[a4paper,margin=0.9in]{geometry}
\usepackage{fancyhdr}
\usepackage{enumitem}
\usepackage{amsmath}
\usepackage{url}
\usepackage{amssymb}
\usepackage{graphicx}
\usepackage{float}
\usepackage{booktabs}
\usepackage[separate-uncertainty=true]{siunitx}
\usepackage[numbers]{natbib} 

\usepackage[ruled,vlined]{algorithm2e}

\usepackage{titlesec}

\titlespacing*{\section}
{0pt}{4ex}{0ex}

\titlespacing*{\subsection}
{0pt}{2ex}{0ex}

\titlespacing*{\subsubsection}
{0pt}{0.6ex}{0.3ex}

\usepackage{caption}
\newcommand{\X}{\Theta}

\usepackage[colorlinks=true,
            linkcolor=blue,
            citecolor=blue,
            urlcolor=blue]{hyperref}

\begin{document}

\title{\LARGE \textbf{Anchored Reversible Jump Sequential Monte Carlo}}

\author{
\textbf{Janna van Assen}\\
\small Department of Industrial Engineering \& Innovation Science\\
\small Eindhoven University of Technology, The Netherlands
\and
\textbf{Max Hinne}\\
\small Department of Machine Learning and Neural Computing\\
\small Donders Institute for Brain, Cognition and Behaviour\\
\small Radboud University, Nijmegen, The Netherlands
}

\maketitle
\begin{abstract}
Reversible Jump Markov Chain Monte Carlo (RJMCMC) is a principled framework for Bayesian model comparison, but its practical use is often limited due to the difficulty of designing between-model proposals that reach regions of high posterior probability. In addition, its inherently sequential nature limits efficient use of modern GPU architectures. We address both challenges by introducing \emph{Anchored Reversible Jump Sequential Monte Carlo}. Our approach combines RJMCMC with Sequential Monte Carlo (SMC) to explore the model space in parallel. Importantly, the particle population enables effective between-model proposals without problem-specific knowledge: particles are used to construct kernel density approximations of the target distributions within each model, which are then used to generate trans-dimensional proposals. Numerical experiments show that the proposed method is computationally efficient and competitive with state-of-the-art RJMCMC approaches, demonstrating its potential for scalable Bayesian model comparison.
\end{abstract}

\section{INTRODUCTION}
Statistical models play a central role in data analysis, providing tools to describe observed patterns and predict future observations. However, in practice, multiple competing models may explain the same dataset. These models may differ in assumptions, functional form, and complexity, potentially leading to different conclusions. Selecting the most appropriate model, known as \emph{model comparison}, is therefore a fundamental problem in statistical inference \citep{burnham_model_2004,piironen_comparison_2017}.

In \emph{Bayesian model comparison}, the model itself is treated as an unknown quantity, and posterior probabilities are assigned to the competing models \citep{kass1995}. For a collection of candidate models $\mathcal{K}$ and observed data $Y$, the posterior probability of model $k$ follows
\begin{equation}
p(k \mid Y) = \frac{p(Y \mid k)\,p(k)} {\sum_{l\in\mathcal{K}} p(Y\mid l)\,p(l)}, 
\end{equation} where $p(k)$ denotes the prior probability of model $k$ and $p(Y\mid k)$ its \emph{marginal likelihood}. The marginal likelihood naturally balances model fit and complexity through an automatic Occam's razor effect, favouring simpler models unless additional complexity is supported by the data \citep{jefferys1992occkam}. Bayesian model comparison therefore provides a principled approach to model selection by quantifying uncertainty over competing models while accounting for the trade-off between model fit and complexity.

Despite its conceptual appeal, Bayesian model comparison is computationally challenging. Posterior model probabilities depend on marginal likelihoods, which are rarely available in closed form and can be costly to approximate for larger models~\citep{llorente_marginal_2023,fourment_19_2020}. To avoid treating models independently, \citet{green1995} introduced \emph{Reversible Jump Markov Chain Monte Carlo} (RJMCMC), a trans-dimensional extension of Metropolis-Hastings that directly explores a joint space of models and estimates posterior model probabilities from the proportion of time spent in each model. However, the practical success of RJMCMC depends critically on the construction of efficient between-model proposal mechanisms that move between regions of high posterior probability across models. Poorly designed proposals often result in low acceptance rates and slow exploration of the model space \citep{hastie2012}. Moreover, effective proposals typically require substantial problem-specific design, making RJMCMC difficult to deploy in practice for complex problems.

To address this challenge, a variety of methods have been proposed to automate proposal construction. \citet{brooks2003} provide a comprehensive overview of reversible jump proposal strategies. Examples of automated approaches include AutoMix \citep{hastie2005}, which employs Gaussian mixture approximations to construct between-model proposals, and more recent transport-based methods \citep{pmlr-v206-davies23a}, which use normalizing flows to learn mappings between posterior distributions. While these methods can substantially improve proposal efficiency, they typically rely on model-specific information obtained through within-model inference. Consequently, applying them to large collections of candidate models remains challenging.

A separate limitation of RJMCMC is its inherently sequential nature. Since inference is performed using a single Markov chain, each sample depends on the previous state, limiting opportunities for parallel computation \citep{liu2001}. In contrast, Sequential Monte Carlo (SMC) methods evolve a population of particles that can be processed in parallel, making them particularly attractive for GPU acceleration \citep{doucet2001,chopin_introduction_2020,hinne_introduction_2025}. This suggests that combining RJMCMC with SMC could substantially improve the computational efficiency of trans-dimensional inference.

\textbf{Contributions.} The main contributions of this work are the following:

\begin{enumerate}[nosep]

\item We introduce adaptive-tempered Reversible Jump Sequential Monte Carlo (RJSMC), a trans-dimensional inference framework that integrates RJMCMC with Sequential Monte Carlo.
\item We introduce \emph{anchored} reversible jump proposals, a novel proposal mechanism that exploits the particle population to construct efficient problem-agnostic trans-dimensional moves.
\item We provide a GPU-accelerated JAX implementation of RJSMC and anchored RJSMC.
\item Across three numerical experiments, we benchmark RJSMC and anchored RJSMC against existing approaches, demonstrating competitive performance and improved computational efficiency in terms of effective sample size per second.
\end{enumerate}

\section{BACKGROUND}

\subsection{Reversible Jump MCMC}
Reversible Jump Markov Chain Monte Carlo (RJMCMC) algorithms were introduced by \citet{green1995} as an extension of  the Metropolis-Hastings algorithm \citep{hastings_monte_1970} to trans-dimensional settings~\citep{green2003,sisson2005,hastie2012}. RJMCMC targets distributions defined on variable-dimensional state spaces of the form
\begin{equation}\label{eq:RJMCMC_state_space}
\X=\bigcup_{k=1}^K{\left(\{k\}\times \X_k\right)},
\end{equation}
where each $k\in\{1,\ldots,K\}$ indexes a model with parameter vector $\theta_k\in\X_k\subseteq\mathbb{R}^{d_k}$, and the parameter space 
dimension $d_k\in \mathbb{N}_{\geq 1}$ may vary across models. In Bayesian model comparison, a natural target distribution is the joint posterior of the model index and parameters given the observed data $Y$:
\begin{equation}\label{RJMCMC target distr}
p(k,\theta_k\mid Y)\propto p(Y\mid\theta_k,k)p(\theta_k\mid k)p(k),
\end{equation}
where $p(\theta_k\mid k)$ and $p(k)$ denote the parameter and model priors, respectively. A Markov chain targeting \eqref{RJMCMC target distr} can then be used to perform inference jointly over models and parameters. To accommodate models of differing dimension, RJMCMC augments the state space with auxiliary variables and dimension-matching transformations. Consequently, RJMCMC can be interpreted as a Metropolis-Hastings algorithm on an augmented space, and inherits its convergence properties under standard assumptions.

Given the current state $(k,\theta_k)$, RJMCMC first proposes a new model index $l$ with probability $j(k,l)$. To construct a move between the corresponding  parameter spaces, auxiliary spaces $\mathcal{U}_k$ and $\mathcal{U}_l$ are introduced such that the dimension-matching condition
\[
\dim(\X_k)+\dim(\mathcal{U}_k)=\dim(\X_l)+\dim(\mathcal{U}_l)
\]
is satisfied. After sampling auxiliary variables $u_k\in\mathcal{U}_k$ from a proposal distribution $g_{k\rightarrow l}$, a diffeomorphism $
h_{k\to l}$
generates the proposed state through 
\begin{equation}
    (\theta_l,u_l)=h_{k\to l}(\theta_k,u_k).
\end{equation}
The dimension-matching condition ensures that $h_{k\to l}$ is invertible, with inverse transformation $h_{l\to k}=h_{k\to l}^{-1}$, thereby enabling reversible jumps between models. The proposed move is accepted with probability
\begin{equation} \label{RJMCMC_acc_ratio}\alpha_{RJ}\bigl((k,\theta_k),(l,\theta_l)\bigr) =  \min\Bigl\{ 1,\, \frac{p(l,\theta_l\mid Y)}{p(k,\theta_k\mid Y)} \frac{j(l,k)}{j(k,l)} \frac{g_{l\to k}(u_l)}{g_{k\to l}(u_k)} \left|J_{h_{k\to l}}(\theta_k,u_k)\right| \Big\}, 
\end{equation}
where $\left|J_{h_{k\to l}}(\theta_k,u_k)\right|$ denotes the absolute determinant of the Jacobian of the transformation $h_{k\to l}$. Implementing RJMCMC therefore requires specifying a jump proposal function $j(k,l)$, auxiliary-variable proposal distributions $g_{k\to l}$, and diffeomorphisms $h_{k\to l}$ for each pair of models $(k,l)$. The efficiency of RJMCMC depends critically on the design of these components. 

While within-model proposals can exploit local posterior information, efficient trans-dimensional proposals require identifying corresponding high-probability regions across models. Without guidance on where to jump, such proposals often land in low-probability regions of the proposed model, leading to frequent rejection and poor mixing across models. Moreover, if transitions between all model pairs are allowed, the number of proposal mechanisms grows quadratically with the number of candidate models, making proposal design increasingly demanding as the model space expands~\citep{brooks2003}. This challenge is further compounded in adaptive model spaces, where proposal mechanisms must be defined for transitions to newly introduced models. Earlier work focused primarily on improving the performance of manually constructed proposals. For example, \citet{AlAwadhi2004Improving} increase proposal acceptance by introducing a secondary Markov chain, while \citet{ehlers_2008_adaptive} and \citet{tian_adaptive_2024} develop adaptive proposal mechanisms that learn from the sampling process. More recently, there is increasing interest in automating the proposal construction. Examples include marginal density estimators \citep{fan_automating_2009}, Gaussian-mixture proposals \citep{hastie2012}, transport-based proposals \citep{pmlr-v206-davies23a}, and approaches that aim to favour transitions towards promising models~\citep{gagnon2021informed}. Although these approaches reduce manual design, they often require substantial preprocessing, such as posterior sampling within individual models, density estimation, or learning model-specific proposal structures. In this work, we pursue a complementary particle-based approach that constructs trans-dimensional proposals directly from the evolving particle population. This further reduces the need for manual tuning while requiring substantially less preprocessing and computation.

\subsection{Sequential Monte Carlo}

Sequential Monte Carlo (SMC) methods approximate target distributions with a population of weighted particles that can be propagated in parallel~\citep{chopin_introduction_2020}. At each iteration, particles are propagated, reweighed according to their consistency with the observations, and resampled to concentrate computational effort in regions of high posterior probability~\citep{Cappe2007}. This population-based framework can improve exploration of complex posterior distributions compared with standard MCMC approaches \citep{del_moral_sequential_2006,hinne_introduction_2025}.

A key challenge in SMC is particle degeneracy: when importance weights become highly uneven, resampling may concentrate the particle population on only a few distinct particles, reducing diversity and exploration efficiency. Tempered SMC mitigates this issue by gradually introducing the likelihood through a sequence of intermediate target distributions \citep{del_moral_sequential_2006},
\begin{equation}
    \pi_n(\theta) \propto p(\theta) p(Y \mid \theta)^{T_n},
\end{equation}
where $p(Y\mid \theta)$ denotes the likelihood of observing the data $Y$ under parameters $\theta$, $p(\theta)$ denotes the prior distribution and $\{T_n\}_{n=0}^N\subseteq [0,1]$ is a non-decreasing temperature schedule. For $T_0=0$, the target distribution equals the prior, whereas $T_N=1$ yields the posterior. At iteration $n$, the target distribution $\pi_n$ is approximated by $M$ weighted particles $\{\theta^{(m)}\}_{m=1}^M$ with normalized weights $\{W^{(m)}\}_{m=1}^M$. To move from $\pi_n$ to $\pi_{n+1}$, the new weights are computed as 
\begin{equation}
    \tilde{W}^{(m)}=\frac{\pi_{n+1}\left( \theta^{(m)}\right)}{\pi_n\left( \theta^{(m)}\right)}=p\left(Y\mid \theta^{(m)}\right)^{\Delta T},
\end{equation} with $\Delta T = T_{n+1} - T_n$, followed by normalization and resampling. The particles are then propagated using a Markov kernel that leaves $\pi_{n+1}$ invariant. In this work, the propagation step employs RJMCMC moves to enable transitions between different models, while within-model updates are performed using standard MCMC. To further avoid particle degeneracy, the temperature is scheduled adaptively based on the \emph{effective sample size} (ESS) of the particle population~\citep{agapiou_importance_2017}. 

While a straightforward trans-dimensional SMC algorithm can be obtained by incorporating RJMCMC moves into the propagation step \cite{BrightMaskell}, we propose several additional components that improve performance, in particular the acceptance rate of trans-dimensional moves. We discuss these modifications next.

\section{PROPOSED METHODS}
\subsection{Reversible jump SMC}
In model comparison, the target distribution is defined over multiple model parameter spaces and may be large, heterogeneous, and highly multimodal. Exploring such a state space with a single Markov chain can be difficult. Reversible jump SMC (RJSMC) combines the trans-dimensional sampling framework of RJMCMC with the population-based exploration of SMC and simultaneously explores parts of the combined model and parameter space. We first introduce a new state representation, followed by an adaptive tempering scheme and a new proposal mechanism.


\textbf{Super-particle Representation}. Rather than storing only the active model state, each RJSMC particle stores states for \emph{all} candidate models together with an active model indicator:
\begin{equation*}
    \left(k^{(m)},\theta_1^{(m)},\dots,\theta_K^{(m)}\right)
\in
\{1,\dots,K\}\times\X_1\times\dots\times\X_K \enspace.    
\end{equation*} Here, $k^{(m)}$ denotes the active model and $\theta_j^{(m)}$ the latent parameter state associated with model $j$. We refer to this augmented particle representation as a super-particle. This representation is related to product-space approaches for Bayesian model selection \citep{carlin_1995_bayesian,Godsill_2001_On}. Its key advantage is that information is retained across candidate models, enabling more informed proposals.

\textbf{Adaptive Temperature Selection}. The super-particle formulation induces two types of importance weights. The standard particle weights,
\begin{equation}\label{particle weight}
    W_{\mathrm{particle}}^{(m)}
    \propto
    p\!\left(Y\mid k^{(m)},\theta_{k^{(m)}}^{(m)}\right)^{\Delta T}\enspace,
\end{equation} are based on the currently active model representation and are used for resampling. Since each super-particle stores states for all candidate models, it also induces representation weights,
\begin{equation}
    W_{\mathrm{repr}}^{(m,k)}
    \propto
    p\!\left(Y\mid k,\theta_k^{(m)}\right)^{\Delta T}.
    \label{representation weights}
\end{equation} Temperature adaptation based solely on the particle ESS can lead to overly aggressive temperature increments. When the active particle population concentrates on a small subset of models, the particle ESS may remain high even though representations of other models have already degenerated. To account for the quality of all model representations, we therefore adapt the temperature using the representation ESS,
\begin{equation}\label{representation ESS}
    M_{\mathrm{eff,repr}}
    =
    \left(\sum_{m=1}^{M}\sum_{k=1}^{K}
    \left(W_{\mathrm{repr}}^{(m,k)}\right)^2\right)^{-1}.\enspace
\end{equation}

Temperature increments are chosen such that $M_{\mathrm{eff,repr}}
\approx
\eta_{\mathrm{repr}} M
$, with $\eta_{\mathrm{repr}}=0.9$ throughout this work. Resampling is subsequently performed using the standard particle weights.

\textbf{Algorithm Description}. RJSMC alternates between adaptive tempering, resampling, and mutation steps. Mutation consists of optional within-model MCMC updates followed by an RJMCMC move. The complete procedure is provided in Appendix~\ref{app:rjsmc}. Since particle updates are independent, RJSMC is naturally parallelizable. The framework is compatible with arbitrary RJMCMC proposal mechanisms, but below we propose a novel problem-agnostic approach.

\subsection{Anchored Reversible Jump Proposals}
We propose a general class of reversible jump proposals that exploits the active and inactive model representations maintained by the RJSMC super-particles. The resulting proposals require no problem-specific mappings or pre-training, but are constructed directly from the current super-particle population using kernel density estimation. Two hyperparameters are introduced: the \emph{anchor weight} $\omega$, which controls the relative influence of active and inactive model representations, and the \emph{grace} parameter $\gamma$, which balances model exploration and exploitation.

\textbf{Independent Parameter-Space Construction}.
To obtain a proposal mechanism applicable to arbitrary model pairs, we make no assumptions about how the parameters of different models are related. Instead, model parameters are treated as independent. For a move from model $k$ to model $l$, we choose $ \mathcal U_k=\X_l$, $\mathcal U_l=\X_k$, and use the diffeomorphism \begin{equation} 
h_{k\to l}(\theta_k,u_k) = (u_k,\theta_k) = (\theta_l,u_l). \label{mydiffeo} 
\end{equation} 
Dimension matching is then automatic, while proposed parameters $\theta_l$ are drawn independently from a proposal density $g_{k\to l}$. Since the Jacobian determinant equals one, the acceptance probability \eqref{RJMCMC_acc_ratio} reduces to 
\begin{equation}\label{BETWEEN_ACC_RATIO}
\alpha=
 \min\left\{ 1, \frac{\pi_T(l,\theta_l)} {\pi_T(k,\theta_k)} \frac{j(l,k)} {j(k,l)} \frac{g_{l\to k}(\theta_k)} {g_{k\to l}(\theta_l)} \right\}, \end{equation} where 
 \begin{equation}
     \pi_T(k,\theta_k) \propto p(Y\mid k,\theta_k)^T p(\theta_k\mid k)\, p(k)
 \end{equation} 
 denotes the current tempered target distribution.

\textbf{Anchored Proposal Distributions}. To propose parameters for model $l$, we exploit the fact that every super-particle already contains a representation $\theta_l$ of that model. The collection $\{\theta_l^{(m)}\}_{m=1}^M$ therefore provides an approximation of the marginal tempered target distribution of model $l$. We construct a weighted Gaussian kernel density estimator
\begin{equation}\label{eq:gaussiankde}
g_{k\to l}(\theta_l)
=
\sum_{m=1}^{M}
\tilde w_l^{(m)}
\,
\mathcal N\!\left(
\theta_l^{(m)},
H_l
\right)\enspace,    
\end{equation} where
\[
\tilde w_l^{(m)}
=
\frac{w_l^{(m)}}{\sum_{j=1}^M w_l^{(j)}},
\qquad
w_l^{(m)}
=
\begin{cases}
\omega, & k^{(m)}=l,\\
1, & \text{otherwise},
\end{cases}
\]
and $H_l$ is the bandwidth parameter determined by the KDE procedure~\citep{Scott1979}. Since particles currently assigned to model $l$ typically provide the most accurate representation of its target distribution, their contribution is upweighted through the anchor weight $\omega\ge1$. As $\omega\to\infty$, only active representations contribute, while smaller values increasingly exploit information from the entire super-particle population. We refer to this approach as \emph{anchored proposals}, as representations of the proposed model act as anchors that guide trans-dimensional jumps toward plausible regions of parameter space.

\textbf{Balancing Exploration and Exploitation}. 
The jump proposal probabilities determine how computational effort is allocated across the model space \citep{gagnon2021informed}. The fraction of particles assigned to model $l$, denoted as $\tfrac{M_l}{M}$, provides a particle-based estimate of its current tempered model probability. We use these estimates to propose destination models according to
\begin{equation}
j(k,l)
=
J_\gamma(l)
=
(1-\gamma)\frac{M_l}{M}
+
\gamma\frac{1}{K}.
\label{grace jump function}
\end{equation}
The exploration-exploitation trade-off induced by the jump proposal distribution $J_\gamma$ is controlled by the grace parameter $\gamma\in[0,1]$. While $\gamma=1$ yields uniform exploration, smaller values increasingly concentrate computational effort on models that currently appear promising, reducing effort spent proposing moves that are unlikely to be accepted.

\begin{algorithm}[h!]
\caption{The A-RJSMC algorithm }\label{algorithmJA}

\textbf{Initialize} $T=0$

\For{\textbf{parallel} $m \in \{1$ \KwTo $M$\}}{
$k^{(m)}\sim p(k)$

\For{$k \in \{1$ \KwTo $K$\}}{
$\theta_{k}^{(m)}\sim p(\theta_{k}\mid k)$}

}

\While{$T < 1$}{

Find $\Delta T$ s.t. $M_{\text{eff,repr}}\approx \eta_{\text{repr}} M$, see (\ref{representation weights}--\ref{representation ESS})

$T \leftarrow \min(1, T + \Delta T)$

$\pi_T \leftarrow p(Y\mid k,\theta_k)^Tp(\theta_k\mid k)p(k)$
    
\For{\textbf{parallel} $m \in \{1$ \KwTo $M$\}}{

        Reweight particles $W_{\text{particle}}^{(m)}$, see \eqref{particle weight}
    }

Resample particles according to $W_{\mathrm{particles}}^{(m)}$

\tcp{Optional MCMC mutation step:}

\For{\textbf{parallel} $m \in \{1$ \KwTo $M$\}}{

    \For{$k \in \{1$ \KwTo $K$\}}{
            $\theta_k^{(m)} \sim q_k(\pi_T)(\theta_k^{(m)},\cdot)$}
    }

\tcp{RJMCMC mutation step:}

\For{$k \in \{ 1$ \KwTo $K$\}}{
Compute KDE $G_{\omega,k}$ as in \eqref{eq:gaussiankde}
}
\For{\textbf{parallel} $m = 1$ \KwTo $M$}{

$(k, \theta_k) := (k^{(m)}, \theta_{k^{(m)}}^{(m)})$

 $l\sim J_\gamma(\cdot)$,  $\theta_l\sim G_{\omega, l}(\cdot)$, $u\sim U[0,1]$

 \If{$u<\alpha_{A}\left((k, \theta_k),(l, \theta_l)\right)$}{
 Accept: $(k^{(m)}, \theta_l^{(m)})\leftarrow \left(l, \theta_l\right)$

 }
}
}

\textbf{Return} particles $\{(k^{(m)}, \theta_1^{(m)},\dots \theta_K^{(m)}\}_{m=1}^M$

\end{algorithm}

\textbf{Algorithm Description}. For a particle currently in active state $(k,\theta_k)$, A-RJSMC first proposes a destination model $l\sim J_\gamma(\cdot)$, and subsequently proposes parameters $\theta_l
\sim G_{l,\omega}(\cdot)$. Under these proposals, the acceptance probability \eqref{BETWEEN_ACC_RATIO} becomes
\begin{equation}\label{JA acc ratio}
\alpha_{\mathrm{A}}((k,\theta_k), (l,\theta_l))
=
\min
\left\{
1,
\frac{\pi_T(l,\theta_l)}
{\pi_T(k,\theta_k)}
\frac{J_\gamma(k)}
{J_\gamma(l)}
\frac{G_{k,\omega}(\theta_k)}
{G_{l,\omega}(\theta_l)}
\right\}. 
\end{equation}
The resulting procedure is shown in Algorithm~\ref{algorithmJA}. An analysis of its complexity is provided in Appendix~\ref{appendix complexity analysis}.

\textbf{Metropolized Independence Interpretation}. The anchored reversible jump kernel is a Metropolized independence sampler, since both $J_\gamma(l)$ and $G_{l,\omega}(\theta_l)$ are independent of the current state $(k,\theta_k)$. Compared to general Metropolis-Hastings algorithms, Metropolized independence samplers enjoy stronger convergence guarantees. Similar ideas have been explored by \citet{South2019SMCSamplers}, who construct independence proposals from particle populations within SMC. Anchored proposals extend this approach to trans-dimensional inference. The full A-RJSMC algorithm is a Metropolized independence sampler only when the optional within-model mutation kernels are omitted or are themselves independence samplers. Besides stronger convergence guarantees, the independence structure enables proposal reuse across particles, making the reversible jump mutation step naturally parallelizable.

\section{EXPERIMENTS}
To evaluate the proposed RJSMC algorithm and the anchored reversible jump proposals, we consider three experiments, as discussed below. 

All experiments were conducted on a single NVIDIA Quadro RTX 6000 GPU (24 GB VRAM) using Python 3.11.15, JAX 0.8.2~\citep{bradbury_jax_2018}, BlackJAX 1.3 ~\citep{cabezas_blackjax_2024}, and BaMoJAX 0.3.12~\citep{hinne_bamojax_2025}. The source code required to reproduce all experiments is available at \url{https://github.com/jannavanassen/RJSMC}.

\subsection{Poisson versus Negative Binomial}
As an illustrative example, we consider the Poisson versus negative binomial model selection problem of \citet{hastie2012}. The availability of an efficient problem-specific reversible jump proposal makes this a useful benchmark for evaluating both the particle-based RJSMC framework and anchored proposals.

The analysis is based on the total-goals dataset comprising 1,040 English Premiership soccer matches from the 2005/06 to 2007/08 seasons. The observed data are goal counts $Y=(Y_1,\ldots,Y_n)$. We compare a Poisson model, $Y_i\sim\mathrm{Poisson}(\lambda)$, with a negative binomial model, $Y_i\sim\mathrm{NegBin}(\lambda,\kappa)$, parametrized such that $\mathbb{E}[Y_i]=\lambda$ and $\mathrm{Var}(Y_i)=\lambda+\kappa\lambda^2$. Thus, the negative binomial model extends the Poisson model by allowing overdispersion through the parameter $\kappa$.

The problem-specific RJMCMC proposal of \citet{hastie2012} exploits the shared mean parameter $\lambda$ by preserving it during trans-dimensional moves and generating the additional dispersion parameter through 
\begin{equation}
    \begin{split}
        u_1&\sim \mathcal N\left(0,\sigma^2\right)\\
        (\lambda_2, \kappa_2)&= h_{1\rightarrow2}(\lambda_1,u_1) = \bigl(\lambda_1,\mu \exp(u_1)\bigr)\enspace. 
    \end{split}
\end{equation} This proposal is highly tailored to the structure of the two models and requires tuning informed by the posterior distribution. In particular, \citet{hastie2012} found that proposals centred at $\mu=1.0$ produced essentially no accepted between-model moves, whereas $\mu=0.015$ yielded effective mixing by proposing values of $\kappa$ within the region of posterior support. Following \citet{hastie2012}, we use $\mu=0.015$, $\sigma=1.5$, and symmetric jump probabilities $j(k,l)=0.5$, and refer to this proposal mechanism as HG-RJMCMC.

\begin{figure}[h!]
    \centering
    \includegraphics[width=0.7\linewidth]{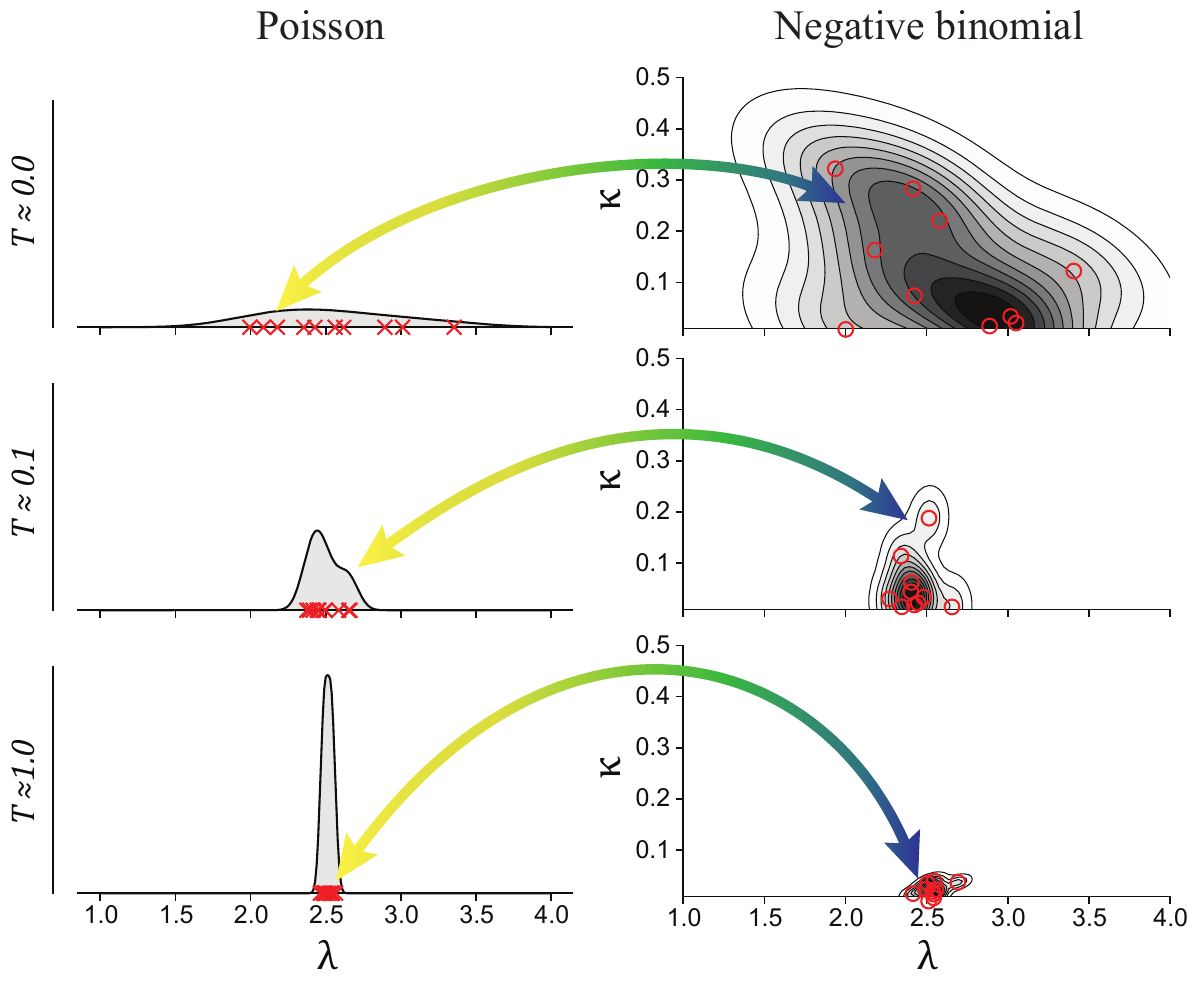}
    \caption{Schematic illustration of the anchored RJSMC proposal mechanism for Poisson versus negative binomial model selection. Each row corresponds to a stage of the adaptive-tempered SMC algorithm. The left and right columns show particles assigned to the Poisson ($k^{(m)}=1$) and negative binomial models, respectively, together with KDEs fitted to their parameters. These KDEs serve as destination-model proposals for trans-dimensional moves. At low temperatures the proposals are diffuse, reflecting the model priors, while as the temperature approaches $T=1$ they converge towards each model's posterior.}
    \label{fig:poisson_vs_negbin}
\end{figure}

We consider six methods for comparison. First, we reproduce the HG-RJMCMC results, using a 50\,000-iteration chain with a 5\,000-iteration burn-in. Second, we envelop the same HG-RJMCMC proposals within a Sequential Monte Carlo (SMC) framework. Finally, we evaluate our proposed RJSMC method with and without additional MCMC rejuvenation steps, and with either HG or anchored (A) reversible jump proposals. This results in four RJSMC variants, each run with 45\,000 particles. The A-RJSMC algorithm is visualized in Fig.~\ref{fig:poisson_vs_negbin}. When MCMC rejuvenation is applied, we use a Gaussian random walk proposal with step size 0.1 and perform two MCMC updates per SMC iteration. For the anchored proposals, we use hyperparameters $\gamma = 0.1$ and $\omega = 100$.

\vspace{5mm}

Table~\ref{tab:full_comparison} summarizes the estimated posterior model probabilities, the maximum potential scale reduction factor (PSRF; $\hat{R}$)~\citep{gelman_rubin_1992_inference} across all parameters, between-model acceptance rates, and computational efficiency metrics for all methods. The reproduced HG-RJMCMC baseline matches the results reported by \citet{hastie2012}. All methods recover comparable estimates of the posterior probability of the Poisson model, with $p(k=1\mid Y)\approx 0.708$. In addition, the maximum PSRF remains close to one for all methods, indicating satisfactory convergence. Note that effective sample size (ESS) in our experiments refers to the number of independent samples in the Monte Carlo approximations~\citep{vehtari_rank-normalization_2021}, not to the SMC weights.

\begin{table*}[h!]
\caption{Comparison of six sampling methods for Bayesian model selection.
Values are reported as mean $\pm$ standard deviation over 10 runs with
different seeds. The maximum PSRF ($\hat{R}$) is taken over the parameters
$\lambda_1$, $\lambda_2$, and $\kappa_2$. The acceptance rate refers to
between-model jump moves only. The between-model acceptance rate could not
be tracked for the HG-RJMCMC in SMC implementation due to limitations of
the software.}
\label{tab:full_comparison}
\begin{center}
\resizebox{\textwidth}{!}{%

\small
\begin{tabular}{
    l
    S[table-format=1.3(1)]
    S[table-format=1.4]
    S[table-format=1.3(1)]
    S[table-format=2.3(3)]
    S[table-format=5.0(3)]
    S[table-format=5.0(3)]
}
\toprule
\textbf{Method}
& {\textbf{$p(k=1\mid Y)$}}
& {\textbf{Max. $\hat{R}$}}
& {\textbf{Acc. Rate}}
& {\textbf{Runtime (s)}}
& {\textbf{ESS}}
& {\textbf{ESS/s}} \\
\midrule

HG-RJMCMC (baseline)
& 0.708(3)
& 1.0002
& 0.584(2)
& 17.72(34)
& 23916(826)
& 1346(45)
\\

HG-RJMCMC in SMC
& 0.707(2)
& 1.0000
& {--}
& 66.26(203)
& 44292(469)
& 669(23)
\\

HG-RJSMC with MCMC
& 0.707(2)
& 1.0000
& 0.625(2)
& 15.76(26)
& 29730(127)
& 1887(33)
\\

HG-RJSMC without MCMC
& 0.707(1)
& 1.0001
& 0.626(2)
& 2.03(1)
& 29076(191)
& 14338(103)
\\

A-RJSMC with MCMC
& 0.708(2)
& 1.0000
& 0.924(1)
& 20.83(25)
& 30568(118)
& 1469(20)
\\

A-RJSMC without MCMC
& 0.707(2)
& 1.0000
& 0.922(2)
& {6.62 $\pm$ 0.00}
& 30054(126)
& 4544(20)
\\

\bottomrule
\end{tabular}}
\end{center}
\end{table*}

RJSMC variants without additional MCMC rejuvenation achieve essentially identical posterior estimates and convergence diagnostics at substantially lower computational cost. This indicates that the reversible jump proposals alone provide sufficient within-model exploration in this low-dimensional setting. Indeed, when $j(k,k)>0$, proposed jumps may remain within the current model $k$, updating parameters either through random-walk perturbations (HG) or resampling from the kernel density estimate (anchored proposals). Consequently, additional MCMC rejuvenation appears unnecessary for this problem.

HG-RJSMC achieves an approximately ten-fold improvement in ESS-per-second compared to the baseline. A-RJSMC achieves the highest acceptance rates overall (approximately 0.92) and, despite the additional cost of constructing kernel density estimates, remains more than three times as efficient as the baseline in terms of ESS per second. A-RJSMC achieves performance comparable to carefully tuned problem-specific HG proposals without relying on manually designed reversible jump mappings, highlighting its potential for generic trans-dimensional inference problems.

\subsection{Variable Selection}
Our second simulation study considers Bayesian variable selection in linear regression, a classical RJMCMC application in which models correspond to different subsets of covariates \citep{green1995,brooks2003}. Efficient problem-specific birth-death proposals have been developed for this setting, providing a strong benchmark for comparison \citep{Forster2012RJM}. The larger model space poses a more challenging trans-dimensional inference problem, providing a test of A-RJSMC in a substantially different setting.

Given a dataset $Y=(X,y)$ consisting of a response vector $y=(y_1,\ldots,y_n)^\top$ and a design matrix $X\in\mathbb{R}^{n\times p}$, each candidate model $M\subseteq\{1,\ldots,p\}$ corresponds to a subset of predictors and is defined by
\begin{equation*}
    y_i=\mu+X_{i,M}\beta_M+\varepsilon_i, \qquad \varepsilon_i\sim\mathcal{N}(0,\sigma^2)\enspace.
\end{equation*} Further details are provided in Appendix~\ref{app:variable_selection}.

As a baseline, we implement a birth-death RJMCMC (BD-RJMCMC) sampler that adds or removes predictors through trans-dimensional moves. We use a whitened parameterization of Zellner's $g$-prior, allowing coefficients associated with newly added predictors to be proposed independently from standard normal distributions, as described in Appendix \ref{appendix birth death rjmcmc implementation}. At each trans-dimensional update, one of the four predictors is selected uniformly and its inclusion status is toggled. The BD-RJMCMC algorithm was run for $21\,000$ iterations, including a burn-in period of $1\,000$ iterations. The A-RJSMC algorithm was run with $20\,000$ particles, $5$ MCMC mutation steps per RJMCMC iteration, $\gamma=0.5$, and $\omega=100$. To ensure a fair comparison, both methods used the same within-model MCMC kernel: an additive random-walk proposal with standard deviation $0.2$. JIT compilation time was recorded separately and excluded from the reported runtimes.

Computational performance and convergence diagnostics are summarized in Table~\ref{tab:performance_metrics}. Both methods accurately recover the target posterior distribution, with correlations between estimated and exact posterior model probabilities exceeding 0.99. Maximum PSRF ($\hat{R}$) values remain close to one throughout, indicating satisfactory convergence. These results show that both methods successfully target the same posterior distribution. While BD-RJMCMC achieves lower runtimes, A-RJSMC produces substantially less correlated samples and consequently attains more than twice the effective sampling efficiency (ESS per second). This improvement stems from the anchored proposal mechanism, which enables direct transitions between all models. In contrast, BD-RJMCMC relies on local birth-death moves that add or remove a single predictor at a time, restricting exploration of the model space and reducing between-model acceptance rates.

\begin{table*}[t!]
\caption{Comparison of BD-RJMCMC (baseline) and A-RJSMC (ours) in the variable selection experiment. Values are reported as mean $\pm$ standard deviation over 10 runs. The maximum PSRF ($\hat{R}$) is taken over the 64 parameters of the super-particle. The acceptance rate refers to between-model jump moves only.}
\label{tab:performance_metrics}

\begin{center}
\resizebox{\textwidth}{!}{%

\small
\begin{tabular}{
    l
    S[table-format=1.3(1)]
    S[table-format=1.3]
    S[table-format=1.3(1)]
    S[table-format=2.1]
    S[table-format=3.1(2)]
    S[table-format=5.0(3)]
    S[table-format=3.1(2)]
}
\toprule

\textbf{Method}
& {\textbf{Corr. exact}}
& {\textbf{Max. $\hat{R}$}}
& {\textbf{Acc. rate}}
& {\textbf{JIT (s)}}
& {\textbf{Runtime (s)}}
& {\textbf{ESS}}
& {\textbf{ESS/s}}
\\

\midrule

BD-RJMCMC
& 0.999(1)
& 1.001
& 0.171(3)
& 3.6
& 28.0(11)
& 1396(171)
& 49.6(61)
\\

A-RJSMC
& 0.997(2)
& 1.003
& 0.467(9)
& 60.0
& 106.6(43)
& 14435(237)
& 136.0(52)
\\

\bottomrule
\end{tabular}

}
\end{center}
\end{table*}

Overall, these results demonstrate that A-RJSMC accurately recovers posterior model probabilities while substantially improving exploration of the model space. Despite requiring no problem-specific proposal design, it outperforms the manually constructed birth-death RJMCMC baseline in terms of effective sampling efficiency, highlighting its potential as a general-purpose approach to trans-dimensional inference.

\subsection{Bayesian Factor Analysis}

\label{sec:bfa}
Finally, we consider Bayesian factor models for multivariate Gaussian data. We revisit the analysis of \citet{Lopes2004BayesianFactorAssessment}, where the objective is to infer the number of latent factors $k$. This problem provides a challenging test of trans-dimensional inference because moves between models involve high-dimensional changes to the factor loading matrix rather than the addition or removal of a single parameter. 

In a $k$-factor model, the covariance structure of the observed variables is represented using a lower-dimensional set of latent factors. Given a dataset $Y=(Y_1,\ldots,Y_n)$ consisting of $n$ independent observations $Y_i\in\mathbb{R}^p$, the model is
\[ Y_i\sim N(0,\Sigma), \qquad \Sigma=\beta\beta^\top+\Lambda, \]
where $\beta\in\mathbb{R}^{p\times k}$ is the factor loading matrix and $\Lambda$ is a positive diagonal matrix. We constrain $\beta$ to be lower triangular with positive diagonal entries, yielding an identifiable parametrization with $p(k+1)-k(k-1)/2$ free parameters. For $i=1,\ldots,6$ and $j=1,\ldots,k$, we assign the priors
\begin{align*}
\beta_{ij}&\sim\mathcal{N}(0,1),\qquad i>j,\\
\beta_{jj}&\sim\mathcal{N}_+(0,1),\\
\lambda_{ii}&\sim\mathrm{IG}(1.1,0.05),
\end{align*}
where $\mathcal{N}_+(0,1)$ denotes a standard normal distribution truncated to the positive real line.

We consider the exchange-rate dataset of \citet[p.~610--618]{west1997bayesian}, consisting of $n=143$ monthly observations of $p=6$ currency exchange rates relative to the British pound from January 1975 to December 1986. Following \citet{Lopes2004BayesianFactorAssessment}, each series is standardized to have zero mean and unit variance prior to inference. We compare models with $k=2$ and $k=3$ latent factors, containing 17 and 21 free parameters, respectively, under a uniform model prior. Using a model-specific RJMCMC algorithm, \citet{Lopes2004BayesianFactorAssessment} originally estimated $p(k=2\mid Y)=0.88$. Subsequently, \citet{South2019SMCSamplers} identified posterior multimodality and generated gold-standard posterior samples using SMC, obtaining $p(k=2\mid Y)=0.894$.

\begin{table*}[h!]
\caption{ Comparison of T-RJMCMC (baseline) and A-RJSMC (ours) in the Bayesian factor analysis experiment. Values are reported as mean $\pm$ standard deviation over 10 runs. The maximum PSRF ($\hat{R}$) is taken over the 38 monitored parameters. The acceptance rate refers to between-model jump moves only.}
\label{tab:rjsmc-results}

\begin{center}
\small
\begin{tabular}{
    l
    r@{\,\(\pm\)\,}l
    c
    r@{\,\(\pm\)\,}l
    r@{\,\(\pm\)\,}l
    r@{\,\(\pm\)\,}l
    r@{\,\(\pm\)\,}l
    r@{\,\(\pm\)\,}l
}
\toprule

\textbf{Method}
& \multicolumn{2}{c}{$p(k=2\mid Y)$}
& \textbf{Max. $\hat{R}$}
& \multicolumn{2}{c}{\textbf{Acc. rate}}
& \multicolumn{2}{c}{\textbf{Runtime (s)}}
& \multicolumn{2}{c}{\textbf{ESS}}
& \multicolumn{2}{c}{\textbf{ESS/s}}
\\

\midrule

T-RJMCMC
& 0.896 & 0.016
& 1.044
& 0.079 & 0.006
& 30933 & 5278
& 771 & 356
& 0.03 & 0.02
\\

A-RJSMC
& 0.897 & 0.023
& 1.027
& 0.067 & 0.013
& 783 & 16
& 14020 & 1982
& 17.62 & 2.40
\\

\bottomrule
\end{tabular}
\end{center}
\end{table*}

\citet{pmlr-v206-davies23a} recently introduced a novel, general-purpose reversible jump approach based on optimal transport proposals constructed from normalizing flows (NFs), which serves as our baseline (T-RJMCMC). We employ the rational quadratic masked autoregression (RQMA) NF. The NF networks are trained on 16\,000 randomly selected draws from the posterior samples of \citet{South2019SMCSamplers}. All chains were run for 100\,000 iterations using the original CPU-based implementation~\citep{pmlr-v206-davies23a}. We run A-RJSMC with 16\,000 particles, $\gamma=0.1$, $\omega = 1000$, and 10 within-model MCMC updates after each RJMCMC move. For both factor models, Hamiltonian Monte Carlo is used as the within-model mutation kernel, with a step size of 0.005 and 20 leapfrog integration steps, enabling efficient exploration of the relatively high-dimensional spaces.

Table~\ref{tab:rjsmc-results} summarizes the performance of the two algorithms across 10 independent runs. Both models show satisfactory convergence according to the potential scale reduction factor (PSRF; $\hat{R}$). The posterior model probabilities match closely with the values found by~\citet{South2019SMCSamplers}. Appendix~\ref{convergence A-RJSMC} shows the evolution of the A-RJSMC model probability estimates over the tempering schedule. Posterior parameter estimates obtained by A-RJSMC also closely match the SMC samples of \citet{South2019SMCSamplers}, indicating successful exploration of the within-model posterior distributions (Appendix~\ref{app:bfa}). Two key differences distinguish the methods. First, T-RJMCMC requires high-quality within-model samples to train normalizing-flow transports before trans-dimensional sampling can begin. Training these neural networks required $218 \pm 85$ seconds on average, excluding the cost of generating the within-model samples. In contrast, A-RJSMC begins trans-dimensional inference immediately, constructing proposals online during sampling. Second, A-RJSMC is designed for massively parallel GPU execution, whereas the T-RJMCMC implementation operates on a CPU. T-RJMCMC required more than six hours to complete a run, while A-RJSMC achieved an effective sample size per second approximately 580 times larger. Appendix~\ref{app:likelihood_evals} provides an implementation-independent comparison based on ESS per model density evaluation. These results highlight the potential of anchored proposals as a general-purpose alternative that integrates proposal construction and posterior inference within a single framework.

\section{CONCLUSION}

This paper introduces Anchored Reversible Jump Sequential Monte Carlo (A-RJSMC) for trans-dimensional inference. A-RJSMC enables parallel trans-dimensional inference without the need for manually tuned model-specific reversible jump mapping.

Across three benchmark problems, A-RJSMC consistently recovered the target posterior distributions and substantially improved sampling efficiency relative to conventional RJMCMC. Moreover, anchored proposals achieved performance comparable to carefully tuned problem-specific reversible jump proposals~\citep{hastie2012} and remained competitive with state-of-the-art transport-based approaches~\citep{pmlr-v206-davies23a}, while requiring minimal problem-specific design and no separate proposal-learning stage. Together, these results move towards automated Bayesian model comparison with efficient trans-dimensional inference across diverse settings.
\newpage

\textbf{Acknowledgements}

This research was funded by the Radboud University ELLIS Excellence Fellowship.

\vspace{5mm}

\bibliographystyle{unsrtnat}

\bibliography{library}

@article{fan_automating_2009,
	title = {Automating and evaluating reversible jump {MCMC} proposal distributions},
	volume = {19},
	copyright = {http://www.springer.com/tdm},
	language = {en},
	number = {4},
	urldate = {2026-09-24},
	journal = {Statistics and Computing},
    author = {Fan, Yanan and Peters, Gareth W. and Sisson, Scott A.},
	year = {2009},
	pages = {409--421},
}

@article{piironen_comparison_2017,
	title = {Comparison of {Bayesian} predictive methods for model selection},
	volume = {27},
	number = {3},
	journal = {Statistics and Computing},
	publisher = {Springer US},
	author = {Piironen, Juho and Vehtari, Aki},
	year = {2017},
	pages = {711--735},
}

@INPROCEEDINGS{BrightMaskell,
  author={Bright, Matt and Maskell, Simon},
  booktitle={2025 IEEE 10th International Workshop on Computational Advances in Multi-Sensor Adaptive Processing (CAMSAP)}, 
  title={Reversible Jump {SMC}}, 
  year={2025},
  volume={},
  number={},
  pages={271-275},
}

@article{Scott1979,
    author = {Scott, David W},
    title = {On optimal and data-based histograms},
    journal = {Biometrika},
    volume = {66},
    number = {3},
    pages = {605-610},
    year = {1979},
    month = {12},
}

@article{vehtari_rank-normalization_2021,
	title = {Rank-{normalization}, {folding}, and {localization}: {An} {improved} {$\hat{R}$} for {assessing} {convergence} of {MCMC} (with {Discussion})},
	volume = {16},
	language = {en},
	number = {2},
	urldate = {2026-10-04},
	journal = {Bayesian Analysis},
	author = {Vehtari, Aki and Gelman, Andrew and Simpson, Daniel and Carpenter, Bob and Bürkner, Paul-Christian},
	year = {2021},
    pages = {667-718}
}

@article{agapiou_importance_2017,
	title = {Importance {sampling}: {Intrinsic} {dimension} and {computational} {cost}},
	volume = {32},
	shorttitle = {Importance {Sampling}},
	language = {en},
	number = {3},
	urldate = {2026-09-25},
	journal = {Statistical Science},
	author = {Agapiou, Sergios and Papaspiliopoulos, Omiros and Sanz-Alonso, Daniel and Stuart, Andrew M.},
	year = {2017},
    pages = {405--431}
}

@article{carlin_1995_bayesian,
author = {Carlin, Bradley P. and Chib, Siddhartha},
title = {{Bayesian Model Choice Via Markov Chain Monte Carlo Methods}},
journal = {Journal of the Royal Statistical Society: Series B (Methodological)},
year = {1995},
volume = {57},
number = {3},
pages = {473--484},
}

@article{Godsill_2001_On,
    author = {Simon J Godsill},
    title = {On the Relationship Between {Markov chain Monte Carlo} Methods for Model Uncertainty},
    journal = {Journal of Computational and Graphical Statistics},
    volume = {10},
    number = {2},
    pages = {230--248},
    year = {2001},
    publisher = {Taylor \& Francis},
    eprint = {https://doi.org/10.1198/10618600152627924}
}

@article{llorente_marginal_2023,
	title = {Marginal likelihood computation for model selection and hypothesis testing: an extensive review},
	volume = {65},
	shorttitle = {Marginal likelihood computation for model selection and hypothesis testing},
	language = {en},
	number = {1},
	urldate = {2024-02-09},
	journal = {SIAM Review},
	author = {Llorente, Fernando and Martino, Luca and Delgado, David and Lopez-Santiago, Javier},
	year = {2023},
	pages = {3--58},
}

@article{fourment_19_2020,
	title = {19 {Dubious} {ways} to {compute} the {marginal} {likelihood} of a {phylogenetic} {tree} {topology}},
	volume = {69},
	number = {2},
	journal = {Systematic Biology},
	author = {Fourment, Mathieu and Magee, Andrew F. and Whidden, Chris and Bilge, Arman and Matsen, Frederick A. and Minin, Vladimir N.},
	year = {2020},
	pages = {209--220},
}

@article{hinne_introduction_2025,
	title = {An introduction to {Sequential} {Monte} {Carlo} for {Bayesian} inference and model comparison—with examples for psychology and behavioral science},
	volume = {57},
	number = {5},
	journal = {Behavior Research Methods},
	author = {Hinne, Max},
	year = {2025},
	pages = {125},
}

@book{chopin_introduction_2020,
	address = {Cham},
	title = {An {Introduction} to {Sequential} {Monte} {Carlo}},
	language = {en},
	urldate = {2023-05-30},
	publisher = {Springer International Publishing},
	author = {Chopin, Nicolas and Papaspiliopoulos, Omiros},
	year = {2020},
}

@ARTICLE{Cappe2007,
  author={Cappe, Olivier and Godsill, Simon J. and Moulines, Eric},
  journal={Proceedings of the IEEE}, 
  title={An Overview of Existing Methods and Recent Advances in {Sequential Monte Carlo}}, 
  year={2007},
  volume={95},
  number={5},
  pages={899-924},
}

@article{del_moral_sequential_2006,
	title = {Sequential {Monte} {Carlo} {Samplers}},
	volume = {68},
	copyright = {https://academic.oup.com/journals/pages/open\_access/funder\_policies/chorus/standard\_publication\_model},
	language = {en},
	number = {3},
	urldate = {2026-09-24},
    journal = {Journal of the Royal Statistical Society: Series B (Statistical Methodology)},
    author = {Del Moral, Pierre and Doucet, Arnaud and Jasra, Ajay},
	year = {2006},
	pages = {411--436},
}

@article{AlAwadhi2004Improving,
    title = {Improving the acceptance rate of reversible jump {MCMC} proposals},
    journal = {Statistics \& Probability Letters},
    volume = {69},
    number = {2},
    pages = {189-198},
    year = {2004},
    author = {Fahimah Al-Awadhi and Merrilee Hurn and Christopher Jennison}
}

@article{ehlers_2008_adaptive,
    author = {Ehlers, Ricardo S. and Brooks, Stephen P.},
    title = {Adaptive Proposal Construction for {Reversible Jump MCMC}},
    journal = {Scandinavian Journal of Statistics},
    volume = {35},
    number = {4},
    pages = {677-690},
    year = {2008}
}

@article{tian_adaptive_2024,
	title = {Adaptive tempered reversible jump algorithm for {Bayesian} curve fitting},
	volume = {40},
	language = {en},
	number = {4},
	urldate = {2026-09-24},
	journal = {Inverse Problems},
	author = {Tian, Zhiyao and Lee, Anthony and Zhou, Shunhua},
	year = {2024},
	pages = {045024}
}

@article{hinne_bamojax_2025,
	title = {Bamojax: {Bayesian} {Modelling} with {JAX}},
	volume = {10},
	copyright = {http://creativecommons.org/licenses/by/4.0/},
	shorttitle = {Bamojax},
	language = {en},
	number = {114},
	urldate = {2025-10-28},
	journal = {Journal of Open Source Software},
	author = {Hinne, Max},
	year = {2025},
	pages = {8642},
}

@misc{cabezas_blackjax_2024,
	title = {{BlackJAX}: {Composable} {Bayesian} inference in {JAX}},
	shorttitle = {{BlackJAX}},
	url = {http://arxiv.org/abs/2402.10797},
	language = {en},
	urldate = {2024-04-23},
	publisher = {arXiv},
	author = {Cabezas, Alberto and Corenflos, Adrien and Lao, Junpeng and Louf, Rémi and Carnec, Antoine and Chaudhari, Kaustubh and Cohn-Gordon, Reuben and Coullon, Jeremie and Deng, Wei and Duffield, Sam and Durán-Martín, Gerardo and Elantkowski, Marcin and Foreman-Mackey, Dan and Gregori, Michele and Iguaran, Carlos and Kumar, Ravin and Lysy, Martin and Murphy, Kevin and Orduz, Juan Camilo and Patel, Karm and Wang, Xi and Zinkov, Rob},
	year = {2024},
}

@misc{bradbury_jax_2018,
	title = {{JAX}: composable transformations of {Python}+{NumPy} programs},
	url = {http://github.com/jax-ml/jax},
	author = {Bradbury, James and Frostig, Roy and Hawkins, Peter and Johnson, Matthew James and Leary, Chris and Maclaurin, Dougal and Necula, George and Paszke, Adam and Van der Plas, Jake and Wanderman-Milne, Skye and Zhang, Qiao},
	year = {2018},
}

@book{burnham_model_2004,
    author    = {Burnham, Kenneth P. and Anderson, David R.},
    title = {Model {Selection} and {Multimodel} {Inference}: {A Practical Information-Theoretic Approach}},
    publisher = {Springer},
     address   = {New York},
     year      = {2002},
     edition   = {Second}
}

@article{kass1995,
  author = {Kass, Robert E. and Wasserman, Larry},
  title = {A Reference {Bayesian} Test for Nested Hypotheses and Its Relationship to the {Schwarz} Criterion},
  journal = {Journal of the American Statistical Association},
  volume = {90},
  number = {431},
  pages = {928--934},
  year = {1995},
}

@article{jefferys1992occkam,
    title={Ockham's Razor and {B}ayesian Analysis},
    author={Jefferys, William H. and Berger, James O.},
    journal={American Scientist},
    volume={80},
    number={1},
    pages={64--72},
    year={1992}
}

@article{green1995,
  title = {Reversible jump {Markov} chain {Monte Carlo} computation and {Bayesian} model determination},
  author = {Green, Peter J.},
  journal = {Biometrika},
  volume = {82},
  number = {4},
  pages = {711--732},
  year = {1995},
}

@incollection{zellner1986,
  author = {Zellner, Arnold},
  title = {On Assessing Prior Distributions and {Bayesian} Regression Analysis with {g}-Prior Distributions},
  booktitle = {Bayesian Inference and Decision Techniques: Essays in Honor of {Bruno de Finetti}},
  editor = {Goel, Prem K. and Zellner, Arnold},
  pages = {233--243},
  year = {1986},
  publisher = {North-Holland},
  address = {Amsterdam}
}

@article{Lopes2004BayesianFactorAssessment,
  author = {Hedibert F. Lopes and Mike West},
  title = {Bayesian Model Assessment in Factor Analysis},
  journal = {Statistica Sinica},
  volume = {14},
  number = {1},
  pages = {41--67},
  year = {2004},
}

@article{South2019SMCSamplers,
  author = {South, Luke F. and Pettitt, Anthony N. and Drovandi, Christopher C.},
  title = {{Sequential Monte Carlo} Samplers with Independent {Markov} Chain {Monte Carlo} Proposals},
  journal = {Bayesian Analysis},
  volume = {14},
  number = {3},
  pages = {773--798},
  year = {2019},
}

@article{gagnon2021informed,
  author = {Gagnon, Philippe},
  title = {Informed reversible jump algorithms},
  journal = {Electronic Journal of Statistics},
  volume = {15},
  number = {2},
  pages = {3951--3995},
  year = {2021},
  publisher = {Institute of Mathematical Statistics},
}

@book{west1997bayesian,
  title = {Bayesian Forecasting and Dynamic Models},
  author = {West, Mike and Harrison, Jeff},
  year = {1997},
  edition = {second},
  publisher = {Springer},
  address = {New York},
  pages = {i--xiv, 1--680}
}

@article{liang2008,
  author = {Liang, Feng and Paulo, Rui and Molina, German and Clyde, Merlise A. and Berger, Jim O.},
  title = {Mixtures of {g} Priors for {Bayesian} Variable Selection},
  journal = {Journal of the American Statistical Association},
  volume = {103},
  number = {481},
  pages = {410--423},
  year = {2008},
}

@article{hastie2012,
  title = {Model choice using reversible jump {Markov} chain {Monte Carlo}},
  volume = {66},
  number = {3},
  pages = {309--338},
  journal = {Statistica Neerlandica},
  author = {Hastie, David I. and Green, Peter J.},
  year = {2012},
}

@article{Forster2012RJM,
  author = {Jonathan J. Forster and Roger C. Gill and Antony M. Overstall},
  title = {Reversible jump methods for generalised linear models and generalised linear mixed models},
  journal = {Statistics and Computing},
  volume = {22},
  number = {1},
  pages = {107--120},
  year = {2012},
  publisher = {Springer},
}

@article{sisson2005,
  author = {Sisson, Scott A.},
  title = {Transdimensional {M}arkov Chains: A Decade of Progress},
  journal = {Journal of the American Statistical Association},
  volume = {100},
  number = {471},
  pages = {1077--1089},
  year = {2005},
  publisher = {Taylor \& Francis}
}

@incollection{green2003,
  author = {Green, Peter J.},
  title = {Trans-dimensional {Markov Chain Monte Carlo}},
  booktitle = {Highly Structured Stochastic Systems},
  editor = {Green, Peter J. and Hjort, Nils L. and Richardson, Sylvia},
  pages = {179--198},
  year = {2003},
  publisher = {Oxford University Press}
}

@InProceedings{pmlr-v206-davies23a,
  title = 	 {Transport Reversible Jump Proposals},
  author =       {Davies, Laurence and Salomone, Robert and Sutton, Matthew and Drovandi, Chris},
  booktitle = 	 {Proceedings of The 26th International Conference on Artificial Intelligence and Statistics},
  pages = 	 {6839--6852},
  year = 	 {2023},
  editor = 	 {Ruiz, Francisco and Dy, Jennifer and van de Meent, Jan-Willem},
  volume = 	 {206},
  series = 	 {Proceedings of Machine Learning Research},
  month = 	 {25--27 Apr},
  publisher =    {PMLR},
}

@article{hastings_monte_1970,
  title = {Monte {Carlo} sampling methods using {Markov} chains and their applications},
  volume = {57},
  number = {1},
  pages = {97--109},
  journal = {Biometrika},
    author = {Hastings, Wilfred K.},
    year = {1970},
}

@book{liu2001,
  author = {Liu, Jun S.},
  title = {{Monte Carlo} Strategies in Scientific Computing},
  publisher = {Springer},
  year = {2001},
  address = {New York},
}

@book{doucet2001,
  author = {Doucet, Arnaud and de Freitas, Nando and Gordon, Neil},
  title = {{Sequential Monte Carlo} Methods in Practice},
  publisher = {Springer},
  year = {2001},
  address = {New York},
}

@phdthesis{hastie2005,
  author = {Hastie, David I.},
  title = {Towards Automatic {Reversible Jump} {Markov Chain Monte Carlo}},
  school = {University of Bristol},
  year = {2005},
  type = {{PhD} thesis}
}

@article{brooks2003,
  author = {Brooks, Stephen P. and Giudici, Paolo and Roberts, Gareth O.},
  title = {Efficient Construction of {Reversible Jump} {Markov} Chain {Monte Carlo} Proposal Distributions},
  journal = {Journal of the Royal Statistical Society: Series B (Statistical Methodology)},
  volume = {65},
  number = {1},
  pages = {3--39},
  year = {2003},
}

@article{gelman_rubin_1992_inference,
    author = {Andrew Gelman and Donald B. Rubin},
    title = {{Inference from Iterative Simulation Using Multiple Sequences}},
    volume = {7},
    journal = {Statistical Science},
    number = {4},
    publisher = {Institute of Mathematical Statistics},
    pages = {457 -- 472},
    year = {1992},
}

\clearpage
\appendix
\titleformat{\section}
{\normalsize\bfseries}
{\thesection}
{1em}
{}

\thispagestyle{empty}

\onecolumn

\section{THE REVERSIBLE JUMP SEQUENTIAL MONTE CARLO ALGORITHM}
\label{app:rjsmc}

The Reversible Jump Sequential Monte Carlo (RJSMC) procedure is summarized in Algorithm~\ref{algorithmRJSMC}. The algorithm takes as input a collection of $K$ candidate Bayesian models for the observed data $Y$. Each model $k$ specifies a prior distribution $p(\theta_k \mid k)$ and a likelihood function $p(Y \mid \theta_k, k)$. In addition, a prior distribution over models $p(k)$, the number of particles $M$, a representation ESS threshold $\eta_{\mathrm{repr}}$, an RJMCMC kernel, and optionally a collection of within-model MCMC kernels $\{q_k\}_{k=1}^K$ must be specified.

Particle mutation is performed using reversible-jump Markov chain Monte Carlo (RJMCMC). Implementing the RJMCMC kernel requires specifying a model-jump proposal distribution $j(k,l)$, auxiliary-variable proposal distributions $g_{k\rightarrow l}$, and diffeomorphisms $h_{k\rightarrow l}$ defining transformations between parameter spaces. Optional within-model MCMC kernels can be used to further explore the posterior distribution of individual models. Such kernels may be particularly beneficial in high-dimensional parameter spaces, where methods such as Hamiltonian Monte Carlo (HMC) can help obtain higher-quality within-model samples before trans-dimensional moves are performed. Here, $q_k(\pi_T)$ denotes a within-model MCMC transition kernel acting on the parameter space of model $k$. The kernel leaves the conditional tempered target distribution of $\theta_k$ given model $k$ invariant.

Algorithm~\ref{algorithmRJSMC} can be combined with arbitrary RJMCMC and within-model MCMC kernels. Existing trans-dimensional samplers may therefore be used directly within the RJSMC framework. The resulting particle population and adaptive resampling stages can lead to substantially more efficient exploration of model and parameter spaces than can be achieved using the underlying MCMC or RJMCMC kernels alone \citep{del_moral_sequential_2006, hinne_introduction_2025}.

\begin{algorithm}[h!]
\caption{The RJSMC algorithm }

\textbf{Initialize} $T=0$

\For{\textbf{parallel} $m \in \{1$ \KwTo $M$\}}{
$k^{(m)}\sim p(k)$

\For{$k \in \{1$ \KwTo $K$\}}{
$\theta_{k}^{(m)}\sim p(\theta_{k}\mid k)$}

}

\While{$T < 1$}{

Find $\Delta T$ s.t. $M_{\text{eff,repr}}\approx \eta_{\text{repr}} M$, see (\ref{representation weights}--\ref{representation ESS})

$T \leftarrow \min(1, T + \Delta T)$

$\pi_T \leftarrow p(Y\mid k,\theta_k)^Tp(\theta_k\mid k)p(k)$
    
\For{\textbf{parallel} $m \in \{1$ \KwTo $M$\}}{

        Reweight particles $W_{\text{particle}}^{(m)}$, see \eqref{particle weight}
    }

Resample particles according to $W_{\mathrm{particles}}^{(m)}$

\tcp{Optional MCMC mutation step:}

\For{\textbf{parallel} $m \in \{1$ \KwTo $M$\}}{

    \For{$k \in \{1$ \KwTo $K$\}}{
            $\theta_k^{(m)} \sim q_k(\pi_T)(\theta_k^{(m)},\cdot)$}
    }

\tcp{RJMCMC mutation step:}

\For{\textbf{parallel} $m = 1$ \KwTo $M$}{

$(k, \theta_k) := (k^{(m)}, \theta_{k^{(m)}}^{(m)})$

 $l\sim j(k, \cdot)$,  $u_k\sim g_{k\to l}(\cdot)$, $(\theta_l,u_l)=h_{k\to l}(\theta_k, u_k)$, $u\sim U[0,1]$

 \If{$u<\alpha_{RJ}\left((k, \theta_k),(l, \theta_l)\right)$}{
 Accept: $(k^{(m)}, \theta_l^{(m)})\leftarrow \left(l, \theta_l\right)$

 }
}
}

\textbf{Return} particles $\{(k^{(m)}, \theta_1^{(m)},\dots \theta_K^{(m)}\}_{m=1}^M$
\label{algorithmRJSMC}

\end{algorithm}

\section{COMPLEXITY ANALYSIS}\label{appendix complexity analysis}

The (serial) time complexity of the standard adaptive-tempered SMC algorithm is $O(MTNC)$, where $M$ is the number of particles, $T$ the number of adaptively determined tempering steps, $N$ the number of mutations per SMC iteration, and $C$ the cost of evaluating one model-specific unnormalized log-posterior density. In a practical implementation on GPU (see the link to our code in the main text), the computation can be parallelized over particles, reducing the effective time complexity to $O(TNC)$. Note that the particle representation has additional benefits on top of reducing the effective time complexity of the algorithm. In an appropriately tempered SMC scheme, several particles can evolve independently, which greatly increases the effective sample size of the approximate posterior, and also avoids getting stuck in case of multimodal posteriors~\cite{hinne_introduction_2025}.

In the case of RJSMC, the model-specific cost $C$ increases as this effectively becomes an evaluation of all unnormalized log-posterior densities in the model collection. Assuming there are $K$ models, the complexity becomes $O(MTNCK)$, or $O(TNCK)$ in an efficient parallel implementation. 

When using RJSMC in conjunction with our anchored proposals, a kernel density is estimated at each tempering step for every particle, by estimating the bandwidth parameter $h$ using Scott's method~\citep{Scott1979}, and subsequently samples are drawn from this density estimate for each trans-dimensional proposal. The additional costs for the kernel construction and bandwidth estimation is $O(MKD)$, with $D=\sum_{k=1}^K d_k$ the total number of model parameters across models. This operation is relatively small and easy to parallelize over $M$, leading to an effective cost of $O(KD)$. The generation of samples from the KDEs however takes a total of $O(TMN^2D)$ additional cost, leading to $O(MTN(KC + MD))$. In our implementation that parallelizes the SMC loop over $M$, the KDE cost remains, so the total (parallel) cost of the anchored RJSMC algorithm is $O(TN(KC + MD)$.

\section{VARIABLE SELECTION MODEL SETUP}
\label{app:variable_selection}
Given a dataset $Y=(X,y)$ consisting of a response vector $y=(y_1,\ldots,y_n)^\top$ and a design matrix $X\in\mathbb{R}^{n\times p}$, each candidate model $M\subseteq\{1,\ldots,p\}$ corresponds to a subset of predictors and is defined by
\[
y_i=\mu+X_{i,M}\beta_M+\varepsilon_i,\qquad i=1,\ldots,n,
\]
where $\mu$ is an intercept, $\beta_M$ is the vector of regression coefficients associated with the predictors in $M$, and the error terms are assumed to be independent with $\varepsilon_i \sim \mathcal{N}(0,\sigma^2)$. To evaluate the proposed inference method in a setting where exact posterior model probabilities are available, we consider Bayesian linear regression with Zellner's $g$-prior \citep{zellner1986}. This prior leads to closed-form marginal likelihoods and Bayes factors \citep{liang2008}, allowing exact posterior model probabilities to be computed for comparison.

For each model $M$, we assign Jeffreys priors to the intercept $\mu$ and residual variance $\sigma^2$,
\[
p(\mu)\propto 1,
\qquad
p(\sigma^2)\propto \frac{1}{\sigma^2},
\]
and Zellner's $g$-prior to the regression coefficients,
\[
\beta_M \mid \sigma^2,M
\sim
\mathcal{N}\!\left(
0,\,
g\sigma^2(X_M^\top X_M)^{-1}
\right),
\]
where $X_M$ contains the predictors included in model $M$. Following \citet{kass1995}, we use the unit-information prior $g=n$ and assume a uniform prior over the $2^p$ candidate models. Under this specification, the Bayes factor of a model $M_k$ with $p_k$ active predictors relative to the null model is
\[
BF_k =
(1+g)^{(n-p_k-1)/2}
\left[1+g(1-R_k^2)\right]^{-(n-1)/2},
\]
where $R_k^2$ denotes the coefficient of determination. Assuming a uniform prior over the $K=2^p$ candidate models, posterior model probabilities are given by
\[
p(M_k\mid Y)
=
\frac{BF_k}{\sum_{j=1}^{K} BF_j}.
\]

We perform our numerical experiment on an artificial data set generated from
\[
y=\mu + X\beta+\varepsilon,
\qquad
\varepsilon\sim\mathcal{N}(0,\sigma^2I),
\]
with
\[
\mu=0,
\qquad
\sigma=1,
\qquad
\beta=(0.2,0.1,0,0),
\qquad
n=100.
\]

This setup contains $p=4$ candidate predictors, yielding $2^4=16$ possible regression models. The regression coefficients are chosen such that posterior mass is distributed across multiple competing models rather than concentrated on a single model. Exact posterior model probabilities are computed from the closed-form Bayes factors above and reported in Table~\ref{tab:posterior_models_exact_vs_estimated}, alongside posterior model probability estimates obtained using A-RJSMC and BD-RJMCMC.

\begin{table*}[t]
\caption{Posterior model probabilities obtained from exact Bayes factor calculations, together with posterior probability estimates from A-RJSMC and BD-RJMCMC. Reported estimates are given as mean $\pm$ standard deviation over 10 independent runs.}
\label{tab:posterior_models_exact_vs_estimated}

\begin{center}
\small
\begin{tabular}{
    c
    l
    S[table-format=1.4]
    S[table-format=1.4(2)]
    S[table-format=1.4(2)]
}
\toprule
\textbf{Model}
& \textbf{Included Covariates}
& {\textbf{Exact $p(k\mid Y)$}}
& {\textbf{A-RJSMC}}
& {\textbf{BD-RJMCMC}} \\
\midrule

0  & None (intercept only)
   & 0.0357 & 0.0379(24) & 0.0341(58) \\

1  & 1
   & 0.2970 & 0.3049(100) & 0.3001(125) \\

2  & 2
   & 0.0142 & 0.0149(10) & 0.0127(19) \\

3  & 1, 2
   & 0.3331 & 0.3238(188) & 0.3284(92) \\

4  & 3
   & 0.0041 & 0.0041(3) & 0.0041(12) \\

5  & 1, 3
   & 0.0447 & 0.0454(27) & 0.0476(22) \\

6  & 2, 3
   & 0.0018 & 0.0020(3) & 0.0018(6) \\

7  & 1, 2, 3
   & 0.0676 & 0.0694(145) & 0.0687(61) \\

8  & 4
   & 0.0062 & 0.0067(6) & 0.0066(12) \\

9  & 1, 4
   & 0.0705 & 0.0727(37) & 0.0702(36) \\

10 & 2, 4
   & 0.0027 & 0.0028(6) & 0.0025(4) \\

11 & 1, 2, 4
   & 0.0935 & 0.0857(173) & 0.0923(62) \\

12 & 3, 4
   & 0.0007 & 0.0007(1) & 0.0007(2) \\

13 & 1, 3, 4
   & 0.0101 & 0.0111(17) & 0.0113(16) \\

14 & 2, 3, 4
   & 0.0003 & 0.0004(1) & 0.0003(2) \\

15 & 1, 2, 3, 4
   & 0.0178 & 0.0176(124) & 0.0186(15) \\

\bottomrule
\end{tabular}
\end{center}
\end{table*}

\section{BIRTH-DEATH RJMCMC IMPLEMENTATION }\label{appendix birth death rjmcmc implementation}

Rather than sampling the regression coefficients $\beta_M$ directly, our BD-RJMCMC implementation uses an equivalent whitened parametrization of Zellner's $g$-prior:
\[
\beta_M=\sqrt{g}\,\sigma\,L\lambda_M,
\]
where $L$ is the Cholesky factor of $(X_M^\top X_M)^{-1}$ and
$\lambda_M\sim\mathcal{N}(0,I)$. This reparameterization expresses the correlated regression coefficients under the $g$-prior in terms of independent standard normal variables. Consequently, during a birth move, the latent parameter associated with a newly introduced predictor can be proposed independently as $\lambda_{M,i}\sim\mathcal{N}(0,1)$. This enables computationally efficient birth proposals by avoiding the need to construct and evaluate proposal distributions that explicitly account for correlations among regression coefficients.

\section{EFFICIENCY OF A-RJSMC AND T-RJMCMC BASELINE}
\label{app:likelihood_evals}

Table~\ref{tab:rjsmc-results} shows how A-RJSMC outperforms the Transport Reversible Jump baseline on the Bayesian factor model in terms of effective sample size per second. However, this result is difficult to interpret due to both the differences in hardware and platforms of the two approaches, as well as the inherent differences between a serial RJMCMC algorithm and a parallel RJSMC approach. We therefore provide an estimate of the number of total density evaluations for both approaches.

\textbf{Transport Reversible Jump}. Davies et al.~\citep{pmlr-v206-davies23a} first approximate the posterior of both models, using NUTS HMC for the 3-factor model and tempered SMC for the 2-factor model. Because both approaches contain adaptive components (for the number of leapfrog integration steps and the number of tempering steps, respectively), we provide a rough estimate. For the 2-factor model, Davies et al.~\citep{pmlr-v206-davies23a} use 16\,000 particles, 2 chains, and a default of 25 MCMC mutations per tempering step. Assuming 10 tempering steps, this leads to a total of $8$M evaluations for the 3-factor model.

The 3-factor model uses 16\,000 draws, plus the PyMC default 1\,000 tuning steps, and again 2 chains. Assuming roughly 50 leapfrog steps for the 3-factor model, this leads to $1.7$M evaluations.

The samples are used to train the RQMA normalizing flows~\citep{pmlr-v206-davies23a}, and are followed by 100\,000 RJMCMC steps. Each step includes a between-model move as with as a within-model step, leading to an additional $200$K evaluations. In total, this results in $9.9$M evaluations, excluding the cost of training the normalizing flows.

\textbf{Anchored Reversible Jump SMC}. Our approach uses 16\,000 particles, 10 mutations each requiring 20 leapfrog integration steps, a maximum of 23 tempering iterations (some runs finished in 22), and evaluations are always done for both the 2-factor and 3-factor model, leading to a total of $147$M evaluations.

With the effective sample sizes given in Table~\ref{tab:rjsmc-results}, this means that TRP results in 78 effective samples per million density evaluations, whereas ARJSMC yields 95 effective samples per million density evaluations. On top of this substantial improvement are the considerations that the normalizing flow training costs was not included the above calculation, nor the gradients of the NUTS sampler, and that even more relative speed improvement follows for the SMC algorithm due to its trivial parallelization.

\section{EVOLUTION OF MODEL PROBABILITIES DURING A-RJSMC}
\label{convergence A-RJSMC}
\begin{figure}[h!]
    \centering
    \includegraphics[width=0.8\linewidth]{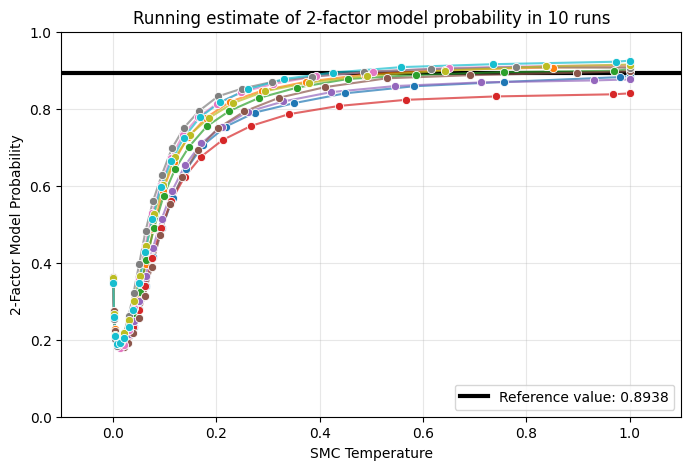}
    \caption{Evolution of the 2-factor model probability across successive temperature updates, shown for 10 independent runs. As the target distribution transitions from the prior distribution to the posterior distribution, the model probability converges towards its posterior probability estimate.}
    \label{fig:exp3 probs}
\end{figure}

Figure~\ref{fig:exp3 probs} shows the evolution of the posterior probability of the two-factor model throughout the tempering schedule. All runs reached the final temperature within 22 or 23 temperature updates, with the model probabilities stabilizing close to the final posterior estimate. In contrast to standard sequential RJMCMC approaches, which require 100,000 sequential iterations in the analyses of \citet{Lopes2004BayesianFactorAssessment} and \citet{pmlr-v206-davies23a}, JA-RJSMC evolves 16,000 particles in parallel over only 22--23 sequential tempering stages. Consequently, the effective sequential depth of the computation is substantially reduced, allowing the method to benefit from parallel hardware and making it suitable for efficient exploration of trans-dimensional posterior distributions.

\newpage

\section{PARAMETER POSTERIORS FOR BAYESIAN FACTOR ANALYSIS}
\label{app:bfa}

\begin{figure}[h!tb]    
    \centering
    \includegraphics[width=0.99\linewidth]{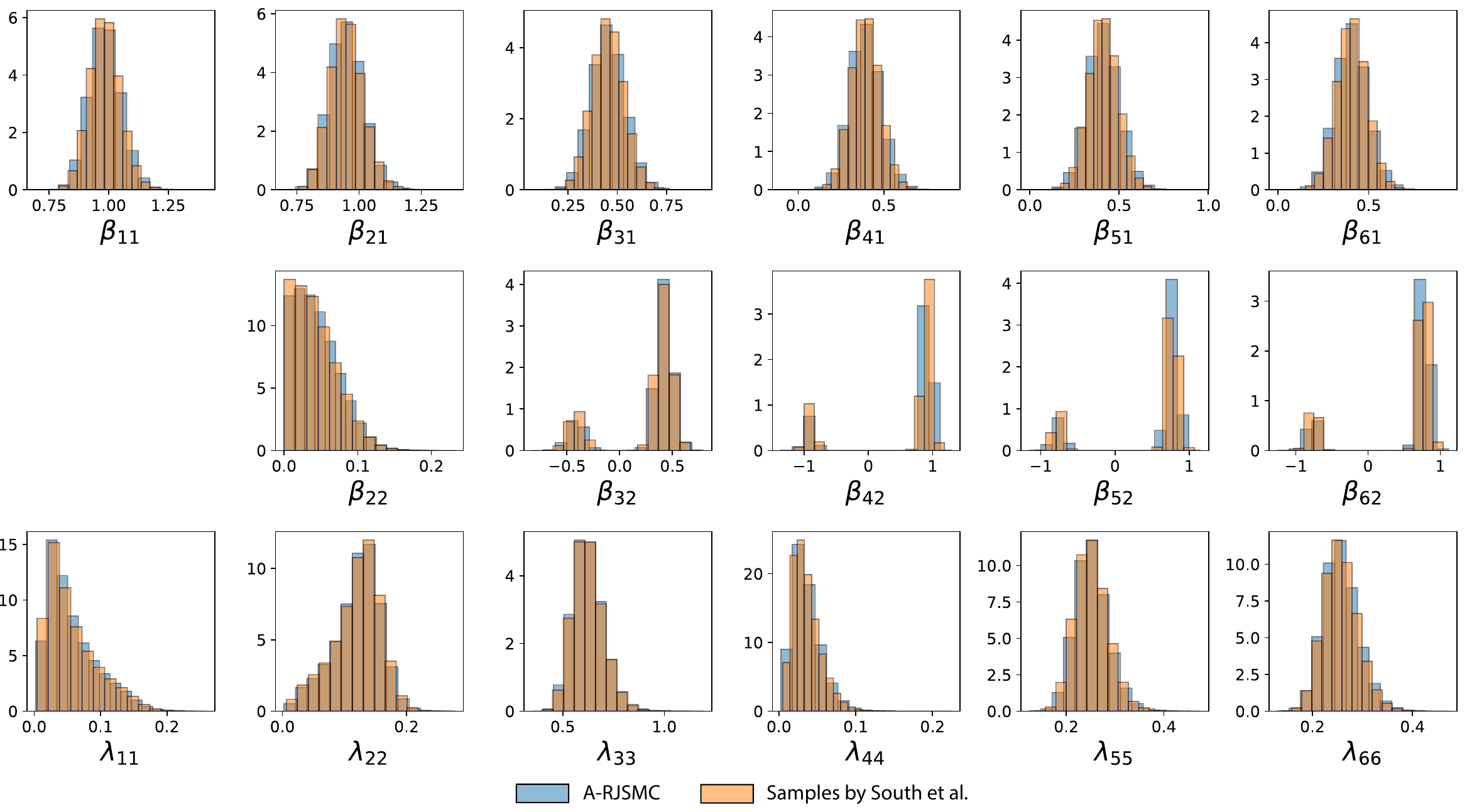}
    \caption{Marginal posterior distributions for all the parameters of the 2-factor model.}      
    \label{fig:b2fa_densities}
\end{figure}

\begin{figure}[h!]    
    \centering
    \includegraphics[width=0.99\linewidth]{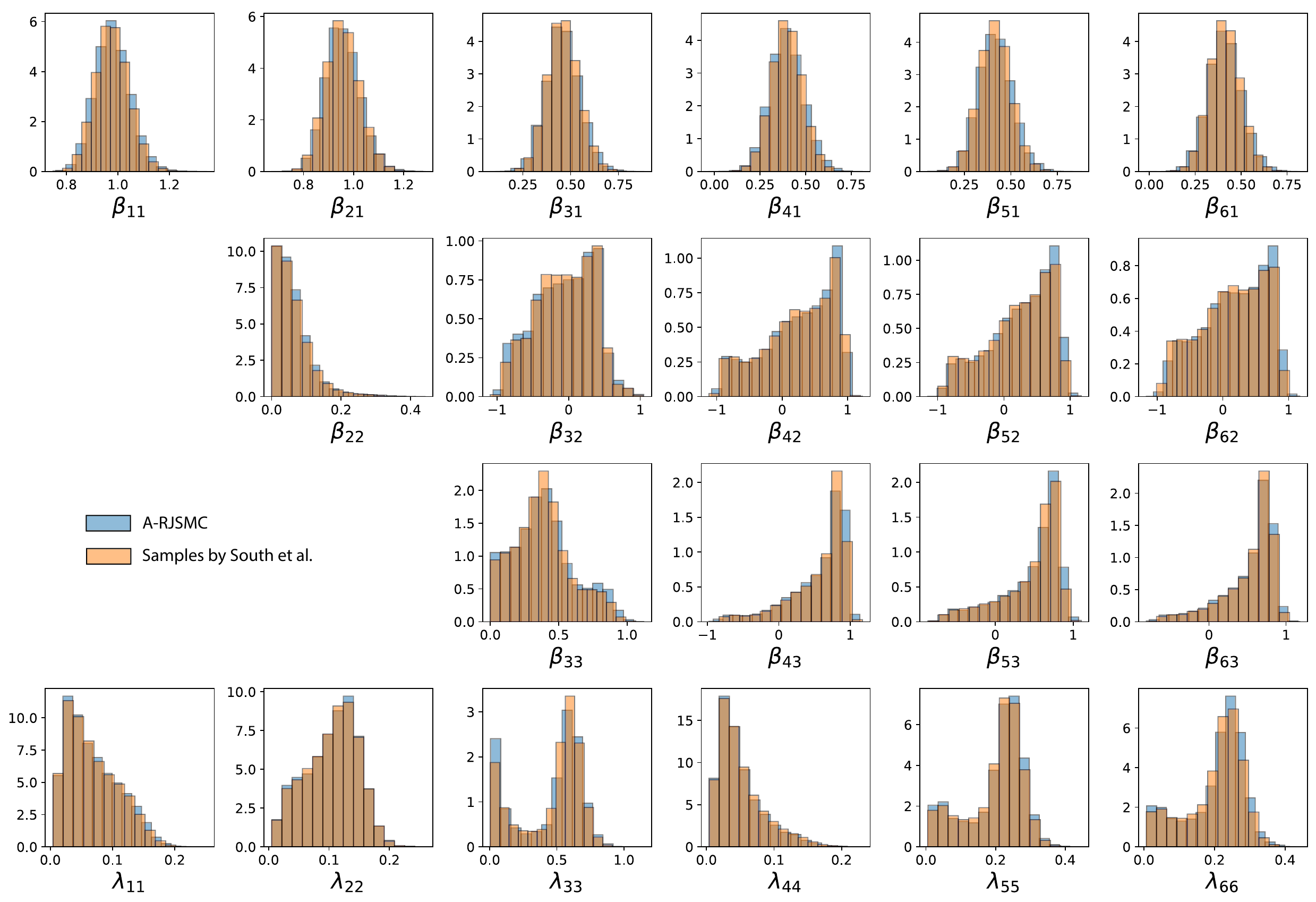}
    \caption{Marginal posterior distributions for all the parameters of the 3-factor model.}       
    \label{fig:b3fa_densities}
\end{figure}

Figures~\ref{fig:b2fa_densities} and~\ref{fig:b3fa_densities} show the posterior densities of the model parameters of both the 2-factor and the 3-factor Bayesian factor models of the experiment in Sec.~\ref{sec:bfa}, for our anchored RJSMC approach as well as the samples obtained by~\citet{South2019SMCSamplers}. Importantly, all densities are virtually identical.

\end{document}